\documentclass[floatfix,prl,aps,twocolumn,showpacs,preprintnumbers,amsmath,amssymb]{revtex4}

\usepackage{graphicx}
\usepackage{graphics}
\usepackage{dcolumn}
\usepackage{bm}

\begin{document}
\title{Collisionally Induced Transport in Periodic Potentials}

\author{H. Ott}
\email{ott@lens.unifi.it}
\author{E. de Mirandes}
\author{F. Ferlaino}
\author{G. Roati}
\author{G. Modugno}
\author{M. Inguscio}

\affiliation{LENS and Dipartimento di Fisica, Universit\`a di
Firenze, and INFM, Via Nello Carrara 1, 50019 Sesto Fiorentino,
Italy}

\date{\today}
\begin{abstract}
We study the transport of ultra cold atoms in a tight optical
lattice. For identical fermions the system is insulating under an
external force while for bosonic atoms it is conducting. This
reflects the different collisional properties of the particles and
reveals the role of inter-particle collisions in establishing a
macroscopic transport in a perfectly periodic potential. Also in
the case of fermions we can induce a transport by creating a
collisional regime through the addition of bosons. We investigate
the transport as a function of the collisional rate and we observe
a transition from a regime in which the mobility increases with
increasing collisional rate to one in which it decreases. We
compare our data with a theoretical model for electron transport
in solids introduced by Esaki and Tsu.
\end{abstract}

\maketitle The motion of particles in periodic potentials is the
underlying process for fundamental transport phenomena like
electric current in metals. Quantum mechanically, particles in a
periodic potential can be described with Bloch states. Without an
external force, the particles can move freely through the
potential and in the case of a non-interacting sample the system
acts like a perfect conductor. Under a constant external force,
the periodic potential is tilted and the new stationary states are
localized Wannier-Stark states \cite{Wannier1960}. In the absence
of interactions the particles cannot change their quantum state
and the latter system behaves like an insulator for DC currents.
Instead, in the presence of interactions collisions can change the
quantum state of the particles and a macroscopic current is
established. At the onset of interactions an increasing
collisional rate is therefore expected to favor a current through
the potential whereas at high collisional rate the current is
hindered by collisions. The latter regime is well known from
solids where scattering with phonons and impurities provide an
extremely large collisional rate and the conductivity decreases
linearly with increasing collisional rates \cite{Ashcroft1976}.
However the limit of low collisonal rate, where the role of
collisions is reversed, is experimentally not accessible in
solids. With the development of semi-conductor superlattices
\cite{Esaki1970}, this regime could be entered and phenomena like
negative electric conductivity could be observed
\cite{Esaki1974,Rauch1998,Agan2001}, but a completely
non-interacting system is not achievable even in superlattices.

In this work we use ultra cold atoms in an optical lattice to
investigate the transport in periodic potentials induced by an
external force starting from the limit of zero interaction. Such
kind of systems have already been used with success to study solid
state phenomena like the Wannier-Stark ladder \cite{Raizen1996} or
Josephson junctions \cite{Cataliotti2001}. Here we take advantage
of the unique possibility of controlling both the scattering
process and the parameters of the lattice to study the transition
from an ideal insulator to a real conductor in a perfectly
periodic potential. The use of indistinguishable fermionic atoms
allows us to create a completely non-interacting system since
inter-particle collisions are forbidden at ultra low temperatures.
On the other hand we can add a scattering mechanism in a
controllable way by using a mixture of fermions and bosons or pure
bosonic samples.

\begin{figure}
\begin{center}
\includegraphics[width=8cm]{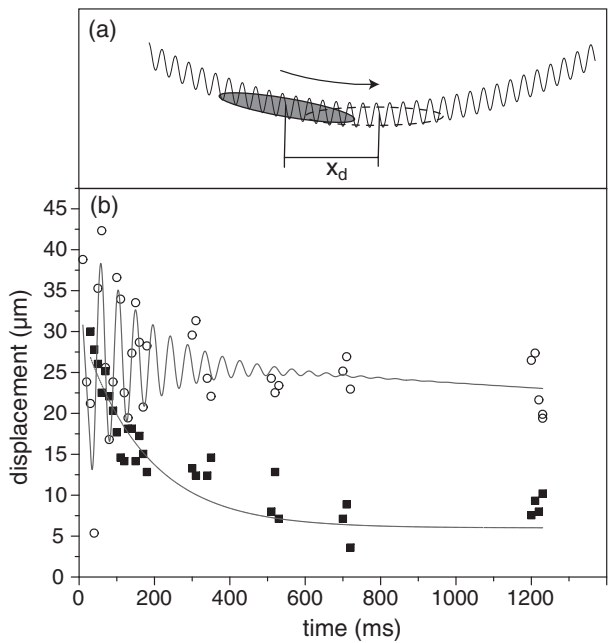} \caption{\label{figure2}(a) Sketch
of the experimental sequence (b) Evolution of the center of mass
(CM) position of a cloud of fermions. A pure fermionic sample
(circles) does not move to the trap center, whereas an identical
sample with an admixture of bosons reveals a current through the
potential (squares). The data are fitted with a sum of an
exponential decay and an initial damped oscillation as described
in the text (continuous lines). The expansion time of the cloud is
$8\,$ms and the lattice height is $s=3$. The temperature and the
atom number of the fermions are $T=300\,$nK and $N=5\times10^4$,
the number of admixed bosons is $N_{\textrm{B}}=1\times10^5$.}
\end{center}
\end{figure}

For the experiments we employ a mixture of ultra cold fermionic
$^{40}$K and bosonic $^{87}$Rb atoms in a magnetic trapping
potential. The spin polarized fermions are sympathetically cooled
by forced evaporation of the bosons \cite{Roati2002,Modugno2003}.
For potassium (rubidium) atoms, the magnetic trap has an axial and
radial oscillation frequency of $\omega_{\mathrm
a}=2\pi\times24(16)\,$s$^{-1}$ and $\omega_{ \mathrm
r}=2\pi\times280(190)\,$s$^{-1}$. The experiments were carried out
with samples at temperatures between $300$ and $400\,$nK. The
number fermions can be varied between $2\times10^4$ and $10^5$
which corresponds to a Fermi temperature of $300-400\,$nK. By
changing the final ramp of the radio frequency evaporation, we can
adjust the number of bosons in the mixture and with a sweep below
the trap bottom it is also possible to remove the bosons. The
temperature of the mixture is always above the critical
temperature for Bose-Einstein condensation. During the last
$500\,$ms of the evaporation ramp, the optical lattice is switched
on adiabatically allowing for thermalization within the lattice.
The lattice is formed by two counter-propagating laser beams and
is aligned in the horizontal plane along the weak axis of the
magnetic trap. Its potential is given by $U(x)=U_0/2(1-\cos(4\pi
x/\lambda))$, where $\lambda=830\,$nm is the wavelength of the
laser. The lattice depth is measured in units of the recoil energy
$U_0=sE_{\textrm r}$ with $E_{\textrm{r}}=\hbar^2k^2/2m$ and
$k=2\pi/\lambda$. In the axial direction, the combined periodic
and harmonic potential has two different kinds of eigenstates that
belong to the first energy band. On both sides of the potential
localized stationary states develop for sufficiently high magnetic
field gradients. These states are the analog to the Wannier-Stark
states in the case of a linear potential and typically extend over
a few ten lattice sites. In the trap center, the states extend
symmetrically around the potential minimum and resemble harmonic
oscillator eigenstates which are modulated by the periodic
potential \cite{dampedoscillation}. The temperature of the samples
is comparable to the recoil energy and is chosen in order to have
a significant occupation of the localized states. The typical
$1/e^2$-radius of the cloud in the direction of the lattice is
$100$\,$\mu$m corresponding to roughly 250 lattice sites. In the
two radial directions, the atoms occupy the radial harmonic
oscillator states. To study the transport of the particles along
the lattice, the magnetic trap is suddenly shifted
\cite{Burger2001} in the direction of the lattice by a fraction of
the extension of the cloud (displacement $x_{\textrm d}$). The
harmonic confinement acts like an external driving force and we
monitor the center of mass (CM) position of the cloud (see
Fig.\,\ref{figure2}a).

The evolution of the CM motion for an initial displacement of
$x_{\textrm d}=35\,\mu$m is shown in Fig.\,\ref{figure2}b. The
open circles show the motion of a pure fermionic sample. The
filled squares show the evolution of the fermionic sample in the
presence of bosons. After an initial damped oscillation, which is
attributed to particles in the harmonic oscillator like states,
the pure fermionic cloud remains displaced in the trapping
potential. This is due to the asymmetric occupation of the
localized states after the displacement which gives rise to an
offset of the cloud with respect to the equilibrium position.
Without collisions, the fermions cannot change the population of
the single particle states. Because Landau-Zener tunnelling into
higher bands is also negligible for our parameters, the offset
cannot vanish and the system is insulating. With an admixture of
bosons however, the fermions rapidly move towards the equilibrium
position \cite{offset}. This macroscopic transport corresponds to
a DC current. To quantify this current, we fit an exponential
decay to the long time tail of the data. For the fermions in the
mixture we find a decay time of $\tau=260\pm30\,$ms, whereas for
the pure fermionic sample the decay time is longer than $5\,$s
which is comparable to the lifetime of the atoms in the optical
potential. This experiment proofs that in a perfect lattice
interactions between the particles are needed to establish a
macroscopic current under an external force.

\begin{figure}
\begin{center}
\includegraphics[width=8cm]{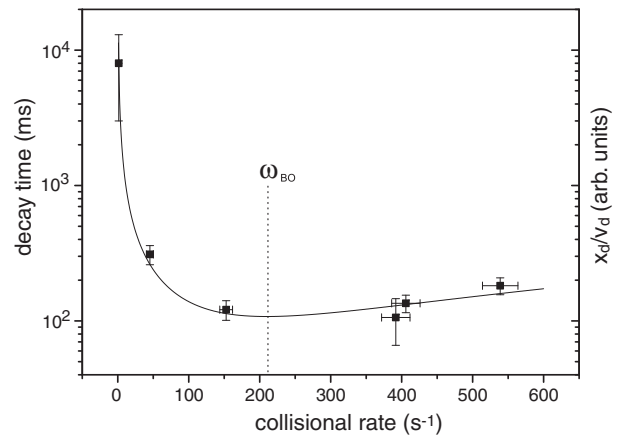}
\caption{\label{figure3}Decay time $\tau$ of a cloud of fermions
in a mixture with bosons in dependence on the collisional rate
(dots). The first data point at zero collisional rate was taken
for a pure fermionic sample. The number of fermions is $N=50\,000$
with a temperature of $350\,$nK. The number of bosons was changed
from 25\,000 to 300\,000 corresponding to a change in the
interspecies collisional rate between $40\,\textrm{s}^{-1}$ and
$550\,\textrm{s}^{-1}$ \cite{collisionalrate}. The lattice height
for the two species was $s_{\textrm{K}}=3$ and
$s_{\textrm{Rb}}=9$, the initial displacement was $x_{\textrm
d}=35\,\mu$m. The solid line is a drift time, calculated from
Eq.\,(\ref{eq1}) for a linear potential with Bloch oscillation
frequency $\omega_{\textrm{BO}}=2\pi\times35\,\textrm{s}^{-1}$
(see text).}
\end{center}
\end{figure}

To investigate the role of collisions in more detail, we have
changed the collisional rate by changing the number of bosons in
the mixture. Fig.\,\ref{figure3} shows the decay time for the
fermionic cloud in a range where the collisional rate was changed
over more than one order of magnitude. Starting from the
collisionless regime, we observe a decrease of the decay time with
increasing collisional rate. This is what one expects if the
collisions assist the hopping between different localized states.
For high collisional rates, the experimental data show a slight
increase of the decay time with increasing collisional rate. In
this regime the number of bosons is much higher than the number of
fermions and the bosons can be regarded as a thermal bath for the
fermions. The fermions exhibit a drift motion and the collisions
with the bosons impede the current through the potential like in a
electric conductor. This change in the mobility of the current
carriers is also known from negative differential conductivity in
semiconductor superlattices \cite{Esaki1970,Esaki1974} (NDC),
which is related to the phenomenon observed here. In the regime of
NDC, the current through the potential decreases when the applied
voltage is increased. This is due to the tighter localization of
the electron wave function which reduces the transition
probability of a hopping event between the localized states
\cite{Tsu1975,Rott1999}. In our experiment, we do not change the
transition probability but the rate of transition inducing
collisions. Despite the different mechanism for NDC only the
product of the transition probability and the collisional rate
determines the final hopping rate. This formal identity allows us
to compare our experimental data with the theoretical model that
was introduced by Esaki and Tsu \cite{Esaki1970} to describe NDC.
The authors calculate the drift velocity of electrons in a
periodic potential under a constant external force. They introduce
a phenomenological scattering rate $\gamma$ and show that the
drift velocity depends on the ratio of the Bloch oscillation
frequency in the linear potential $\omega_{\textrm{BO}}$ and the
scattering rate $\gamma$:
\begin{equation}\label{eq1}
v_{\textrm{d}}=v_0/4\frac{\omega_{\textrm{BO}}/\gamma}{1+\left(\omega_{\textrm{BO}}/\gamma\right)^2},
\end{equation}
with $v_0=\lambda\Delta E/\hbar$ being the tunnelling speed
through the potential and $\Delta E$ being the width of the first
band. A direct adaptation of the above equation to our dynamics is
rather complicated because we have a spatially varying Bloch
oscillation frequency \cite{Blochfrequency} and an inhomogeneous
system. However, we can compare the initial velocity $v_i$ of the
center of mass that we observe in the experiment with the drift
velocity calculated from Eq.\,(\ref{eq1}) for a uniform system in
a linear potential. To determine the Bloch oscillation frequency
in this potential, we take the force that initially acts on the
center of mass after the displacement and we identify the
scattering rate $\gamma$ with the average collisional rate between
the fermions and the bosons. Because the initial velocity of the
center of mass is connected to the decay time by $\tau=x_d/v_i$ we
can also compare the decay time $\tau$ with the inverse of the
drift velocity $v_d$. The result is shown in Fig.\,\ref{figure3},
where we have computed the solid line leaving $v_0$ in (\ref{eq1})
as a free fitting parameter. In spite of the simplification the
model reproduces well both the drop of the decay time at low
collisional rates and its slight increase for high collisional
rates. This supports the phenomenological interpretation given
above.

Because we have to deal with inter-particle collisions, our
damping mechanism is different from that in superlattices and we
have to ask if the assumptions made in Ref.\,\cite{Esaki1970} to
derive expression (\ref{eq1}) are fully valid in our system. The
phenomenological scattering rate $\gamma$ describes dissipative
scattering processes, where the electrons can arbitrarily exchange
energy and momentum with an external thermal bath. In our system,
no energy exchange with the lattice is possible because the
lattice is free of impurities or excitations and momentum can only
be transferred to the lattice in multiples of the Bragg momentum
via umklapp scattering processes. As mentioned above, in the case
of a mixture of fermions and bosons, the latter might be regarded
as a thermal bath for high atom numbers and a dissipative
scattering channel exists. For small numbers of admixed bosons and
for pure bosonic samples, where the assumption of having a thermal
bath is questionable, we find the same phenomenology. This
indicates that also in this case a dissipative mechanism is still
present, possibly related to the coupling to the two radial
degrees of freedom.

\begin{figure}
\begin{center}
\includegraphics[width=8cm]{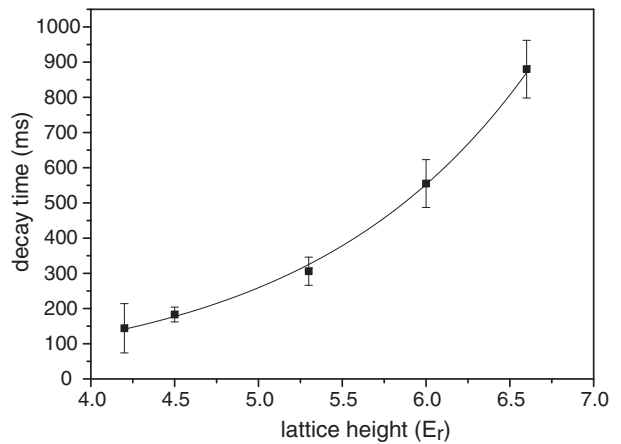}
\caption{\label{figure4}Decay time of a cloud of bosons for
different lattice heights. The initial displacement was
$x_{\textrm d}=10\,\mu m$. The continuous line is an exponential
fit to the data. The exponent is given by $e^{-s/1.6}$.}
\end{center}
\end{figure}

We have also studied how the transport of the atoms is affected by
different lattice potentials which can be easily tuned by changing
the laser intensity. In experiments carried out with thermal
bosons, we have measured the dependence of the decay time on the
lattice height, such as shown in Fig.\,\ref{figure4}. The data
show a rapid increase of the decay time with increasing lattice
height. This can be explained with a reduction of the tunnelling
probability between neighboring lattice sites with increasing
lattice height. In expression (\ref{eq1}) the tunnelling speed
appears as a scaling factor for the drift velocity. Even if one
takes into account a spatially varying Bloch oscillation frequency
$\omega_{\textrm BO}(x)$ and an inhomogeneous scattering rate
$\gamma(x)$, the role of $v_0$ does not change. Thus, we can write
for each single particle a differential equation of the form
$\dot{x}/v_0=f(\omega_{\textrm BO}(x),\gamma(x))$, whose solution
scales in the time domain with $v_0$. Consequently, also the
behavior of the center of mass scales with $v_0$ and the decay
time must be proportional to the inverse of the bandwidth. For a
sinusoidal potential, the bandwidth can be expressed in terms of
Mathieu functions and we find that for $s<10$ the bandwidth is
well described (the maximum error is smaller than 10 percent) with
an exponential drop of the form $\Delta E=E_{\textrm
r}e^{-s/3.8}$. One therefore expects an exponential increase of
the decay time with increasing lattice height. The exponential fit
in Fig.\,\ref{figure4} demonstrates well that this dependence is
accomplished. We find a numerical value for the factor in the
exponent of the fit of $1.6$. For other experimental data sets
with different temperatures and initial displacements we derive
values ranging from $1.5$ to $4.5$.

We can now identify two crucial processes that are needed for the
macroscopic transport through a periodic potential in the presence
of an external force. The first one is the tunnelling from one
lattice site to the next one. However, the coherent nature of the
tunnelling process leads to a localization of the particle.
Therefore an additional dissipative process is needed to destroy
the localization of the particle wave function. If one of these
two mechanisms is missing, the system is insulating, as we observe
it for non-interacting fermions and in the limit of deep lattices.

In conclusion we have demonstrated that under an external force
the macroscopic transport of particles in a perfect periodic
potential requires an interaction between the particles. Indeed,
in the non-interacting case, we find an insulating behavior of the
system. In the interacting regime, we observe a transport, which
is mediated by collisions. We have investigated the dependence of
the transport velocity on the collisional rate and on the lattice
height. A comparison with a semiclassical band model
\cite{Esaki1970} introduced for electrons in superlattices reveals
a good qualitative agreement although the microscopic dissipative
mechanism is different. In this first study of the transport of
fermionic atoms in optical lattices the interaction was provided
by an admixture of bosons. In the future it will be interesting to
prepare the fermions in two different spin states and to extend
the study to the regime of high quantum degeneracy. An even more
flexible way of tuning the interaction can be achieved by using a
Feshbach resonance. Such a system of interacting fermions would be
interesting for possible studies of the influence of fermionic
superfluidity on the transport along a periodic potential.

We want to thank G. La Rocca for illuminating discussions and T.
Ban for contributions to the experiment. This work was supported
by MIUR, by EU under Contract. No. HPRICT1999-00111, and by INFM,
PRA "Photonmatter". H.\,O. was supported by EU with a Marie Curie
Individual Fellowship under Contract No. HPMF-CT-2002-01958.


\begin{references}
\bibitem{dampedoscillation} A detailed study of the single particle eigenstates
of the combined potential will be addressed in a forthcoming
paper.
\bibitem{offset}
Due to a heating of the radial degrees of freedom the axial
extension of the cloud in the mixed system increases. Because an
existing anharmonicity of the trapping potential shifts the center
of mass for increasing temperatures, the cloud remains slightly
above the equilibrium position.
\bibitem{Blochfrequency}
The energy difference $\delta E(x)$ between two adjacent wells
grows linearly with the distance from the trapping center $\delta
E = m\omega_a^2x\lambda /2$. The Bloch oscillation frequency,
given by $\hbar\omega_{\textrm{BO}}=\delta E(x)$ therefore
linearly depends on the position.
\end{references}
\end{document}